# Towards a Sociology of Sociology: Inequality, Elitism, and Prestige in the Sociological Enterprise From 1970 to the Present

Gavin G. Cook

*University of Macau and Harvard University*

[Word count: 9,579]

**Acknowledgments**

- I wish to thank Dalton Conley, Brandon Stewart, Miguel Centeno, Yu Xie, Xi Song, Hyunjoon Park, Emily Hannum, and Arthur Sakamoto for valuable comments while preparing this manuscript. I also thank Xudong Guo for valuable research assistance.

**Key Words**

- Stratification, elitism, sociology, inequality

**Corresponding author**

- Gavin G. Cook

- gcook@fas.harvard.edu, +1-310-383-6580

**ABSTRACT**

There is a science of science and an informal economics of economics, but there is not a cohesive sociology of sociology. We turn the central findings and theoretical lenses of the sociological tradition and the sociological study of stratification inward on sociology itself to investigate how sociology has changed since the 1970s. We link two bibliometric databases to trace diachronic relationships between PhD training and publication outcomes, both of which are understudied in the science of science and sociology of science. All of sociology's top 3 journals remained biased against alum of less prestigious PhD programs, and while most forms of bias in elite sociological publishing have ameliorated over time, the house bias of the American Journal of Sociology in favor PhD alumnae of UChicago has intensified.

# INTRODUCTION

In the words of one economist, "navel-gazing has been a favorite activity of economists (and other academics) for at least sixty years" (Hamermesh 2013, 162). Sociologists are no stranger to looking inward at their own field, but they often avoid gazing too intently at the ugly imperfections of the sociological establishment.

This is not surprising. The first and most fundamental theory of the sociology of science might be termed the Imperfect Science Hypothesis. It argues that science is ostensibly meritocratic and objective, but, as any scientist can attest and as many sociologists of science have argued, scientists are only human, and humans, as all humans know, are imperfect. Science is, therefore, as imperfect as the scientists that practice it (Merton 1974; Latour and Woolgar 1986; Xie and Shauman 2009). This assumption undergirds the entire enterprise of the sociology of science, and almost all empirical work on the topic, both qualitative and quantitative, has proven this assumption to be true.

A famous but fraught example of science's imperfections may be found in the relationship between PhD prestige, journal publications, and job placement. It is broadly assumed by most academics that journal articles, ideally highly-cited and in prestigious journals, are the common currency of academic advancement and determine job placement. Sociologists of science have long argued that this is not the case; instead, they contend the prestige of one's PhD-granting institution and advisor determine an academic's professional success (Long 1978; Long et al. 1979; Allison and Long 1990). Indeed, some have even argued that the presumed causal relationship between productivity and professional achievement may actually run in reverse, with the prestige of an academic's professional context determining productivity (Long and McGinnis 1981). This is a scathing refutation of the Mertonian norm of science as a

'disinterested search for truth' (Merton 1974, 323). This is on its own is distressing enough, but it would only be worse if the publication process were also biased.

The question of publication bias has attracted very little attention. This is perhaps because it does not behoove any professional to bite the hand that feeds him or her, and it was very difficult until very recently to attain access to the data that would make the tentative evaluation of bias in journals possible. Any exploration of bias in the publication process would be particularly relevant for sociologists because the distribution of prestige accorded to sociology's major journals is extraordinarily unequal. While sociologists may not agree on the horizontal boundaries of the field's core and many frontiers, they certainly agree upon the general outlines of how the field stratifies vertically: the AJS, ASR, and Social Forces are at the top of sociology, and journals become more specific and less prestigious from there on down. A publication in any of sociology's top 3 journals can be seen as an argument-ender between feuding camps of professors from different subfields in hiring committees. Because of their primacy in the hiring process, publications in sociology's top 3 journals are a valuable but understudied source of data for studying the sociological enterprise. It has long been noted that the hierarchy between sociology journals emerges no matter how one measures it or on what time horizon one analyzes it (Jacobs 2016). Even in the 1990s, sociologists had observed that the competition to publish in top journals grew more intense year after year, making "journals the unequivocal courts of first instance for professional success" (Abbott 1999, 190).

If those 3 journals form the narrow peak of sociology, the field's wide base is the dissertation. Dissertations are read by very few and are generally not seen as prestigious. They are a perfunctory exercise that every sociologist must complete and may also turn into journal or monograph-length publications. From base to peak, then: every academic sociologist must complete a dissertation, but not every academic sociologist publishes journals, for some attrit from the academy, and far fewer academic sociologists ever publish in sociology's top 3 journals.

To understand who makes it to sociology's peak, we must understand how the peak of sociology relates to the base. This was not possible until very recently. To do so, I link a database containing every sociology thesis published since 1970 to another bibliometric database that contains almost every journal article published in sociology. This allows us to explore how PhD prestige may influence publication outcomes. While I do not and cannot claim to have comprehensively summarized every element of the American sociological establishment, this dataset provides a robust descriptive overview of elitism in American sociology.

There is very little empirical inquiry in sociology on the sociological enterprise, and sociology is overdue for a temperature-taking. Following the scientists of science (Fortunato et al. 2018) and a tradition that we name the economics of economics (Heckman and Moktan 2020), we will turn one of our field's grandest traditions, specifically the sociological study of inequality, on sociology itself. There has been a scattered flurry of profoundly insightful sociological work on the sociological enterprise over the past sixty years, particularly in the 1970s (Shamblin 1970; Yoels 1971). This pioneering work unfortunately never condensed into a subfield with legs of its own the way it has in economics or other social sciences. While this is unfortunate, it is not surprising because studying sociology poses a few major difficulties. Chief among these difficulties is that sociology is a notoriously nebulous field with no formal core. Sociology is so low in consensus that Thomas Kuhn's theory of the scientific paradigm, one of the most well-known theories in all of science studies, was inspired in part by watching sociologists argue amongst themselves at the Institute of Advanced Study (Kuhn 2012 [1962], 21). Thirty years after Kuhn's work, Cole summarized the state of sociology in the 90s with a single, pithy sentence: "There seems to be no sociological work that the great majority of the community will regard as both true and important" (Cole 1994, 134). In the core-frontier framework of Cole (1983, 1992), it is very difficult to pinpoint what the core of sociology contains, and it is more difficult still to delineate where the boundaries of the frontiers of sociology lie. Early and more recent empirical explorations of this topic have confirmed that sociology is among the lowest-consensus fields of study in the academy (Merton 1974; Cook and Xie 2023).

  We devote special focus to elitism in the form of a preference for alumni of high-ranking PhD programs. There are certainly other vectors of bias in academia, the most prominent of which follow the societal fissures of race and gender: the authors of journal articles are overwhelmingly male (Son and Bell 2022), as are the editors of journals, and male editors tend to self-publish in their own journals at higher rates than do women (Liu et al. 2023). Additionally, women are less likely to receive credit in the form of authorship on a paper for their work (Ross et al. 2022), and, in physics, papers authored by women are cited less (Teich et al. 2022). While gendered and racialized bias operate in sociology as well, critiques of elitism occupy a special place in the sociology of science and in the nascent sociology of sociology (Shamblin 1970; Long et al. 1979; Allison and Long 1990). We aim to revisit the theoretical focus on prestige that was emblematic of these two traditions. While there have been many pioneering works on elitism published recently (Khan 2012; Khan 2013; Sherman 2017; Cousin et al. 2018; Keister et al. 2022), the sociological literature on elitism is generally thinner than the literatures on the vectors of inequality that correspond to identity categories, such as race, gender, and migration status, and this literature contains great and untapped theoretical value for analyzing inequality in sociology itself.

The follow analysis is comprised of two parts. We first analyze the interaction between prestige and publication outcomes in sociology's three leading journals: the AJS, the ASR, and Social Forces. Drawing upon methods and theory from the of sociological study of inequality, we find that articles penned by alumnae of prestigious PhD programs are over-published but under-cited in top sociology journals and that this bias has been more or less constant for decades. This bias in favor of prestige implies a halo effect (Thorndike 1920) of prestige on publication outcomes and is a violation of the meritocratic norms of science in general and sociology specifically. We then examine one of the testiest topics of water-cooler discussion in sociology departments across America: does the close connection between the UChicago sociology department and the AJS invite nepotism? We find that house bias towards UNC-CH alum in Social Forces, a situation roughly analogous to that of UChicago and the AJS, was strong in the past but has since dwindled to almost nothing in the present. By contrast, house bias favoring UChicago PhD alum in the AJS has generally increased over the past two decades. Sociology, then, is unfortunately unequal, which has profound consequences for a field with such a profound focus on the empirical and theoretical study of inequality. More broadly, we hope to make a case for the sociology of sociology and invite others to analyze sociology with the methods of the sociologist.

**LITERATURE REVIEW**

**WHAT PUBLICATION BIAS IS AND IS NOT**

The terminology surrounding biases and effects in the social sciences is inherently confusing. A central issue regarding the definitions of common types of bias is that their phrasing rarely indicates the direction of the action or activity that may be biased. Publication bias is a first-rate example of this phenomenon. In most cases, publication bias is generally understood to refer to a bias regarding what types of results are submitted for publication (van Aert, Wicherts, and van Assen 2019; Leichsenring et al. 2017; Franco, Malhotra, and Simonovits 2014). Publication bias understood thusly is often diagnosed with funnel plots, where published effect sizes are plotted in tandem with measures of precision (Mavridis and Salanti 2014). Asymmetry in funnel plots is taken to indicate publication bias, and these plots are frequently used in meta-analyses in biomedical research (Sterne et al. 2011). Sociologists have used other statistical tests to examine the extent of this framing of publication in our own field (Gerber and Malhotra 2008).

Still more confusing is that many definitions of publication bias do not differentiate between bias in authors submitting papers for publication and in journal editors' reception of these publications (Eastbrook et al. 1991; Song, Hooper, and Loke 2013). These discrete stages are instead bundled together and treated as one. The almost universal disdain for null findings that is reported across the publication bias literature is then treated as stemming from either element. We focus not on publication submission bias but what might be termed publication reception bias.

**REFLEXIVE META-FIELDS AND THE ECONOMICS OF ECONOMICS**

The landscape of academia is peppered with a mixture of both formally and informally defined subfields that train the methods of a given field on the field itself. We refer to these subfields as recursive meta-fields. Before we propose a sociology of sociology, we will first summarize extant work in other recursive meta-fields. The most famous of these is the science of science (Fortunato et al. 2018), which purports to use the scientific method to science but informally amounts to the quantitative study of science at scale. The scientists of science have produced pioneering work on the influence of team structure on science (Jones, Wuchty, and Uzzi 2008; Adams 2013), the lifecycle of citations (Yin and Wang 2017), and productivity across the career (Liu et al. 2021). While the science of science is a formal subfield, the other recursive meta-fields are generally less formal. These less formal fields include the loose body of work that may be termed the psychology of psychology (Blasi et al. 2022; Atari, Henrich, and Schulz 2025; Kroupin et al. 2025), and historiography, which may be read as a history of history (Cheng 2012). We must look elsewhere, then, for a model from which to build a sociology of sociology.

Economics is the only social science with an active if informally defined reflexive meta-field of its own. While the discussion of why this may be the case is beyond the scope of this article, the bibliometric of the economics of economics is substantial. Though the authors who publish in its pages do not describe their work with the term, the Journal of Economic Literature is dedicated to what may be termed an economics of economics. The sheer size and relative cohesion of the recursive meta-field of economics makes it a very well-defined and stable departure point for our sociology of sociology. While sociologists have devoted very little empirical attention to sociology's top journals, economists have analyzed publishing trends in their field's most prestigious journals in great detail and at great length. Economics, a larger field by far than sociology, has five top journals to sociology's three. As is the case for publications in sociology's top three journals, a publication in a top five economics journal is crucial to professional advancement in economics. Publishing in the top five journals, however, has only grown more difficult. Top 5 economics journal acceptance rates from dropped from 15% in 1976 to 6% in 2012 as annual submission rates doubled from 1990 to 2011 (Card and DellaVigna 2013).

Any bias in this process would taint the putative meritocracy of economics, and, at first blush, it appears that there is bias in this process. Of particular concern is that the world of economics has grown more nepotistic in general; the central node in economics coauthorship networks comprised 15% of all connections in the 1970s but 40% of all connections in the 2000s (Goyal et al. 2006). This nepotism is especially evident in top five journals, which show favoritism to scholars who are connected to the members of their editorial boards (Heckman and Moktan 2020, 422). One survey of connections between journal editors and the scholars who publish in said journals estimates that 43% of the 1,620 articles in 4 of the top 5 journals published between 2000 and 20006 were authored by a scholar with a connection to a member of the editorial board. Specifically, the former doctoral students of editors wrote 15% of the articles and former colleagues a staggering 29% of the articles in these journals (Colussi 2018). While some economics have argued that nepotism may be used to identify high-impact papers more efficiently (Laband and Piette 1994), far more economists have instead explored how nepotism warps the metaphorical marketplace of ideas in economics. Given the relative dearth of work on this subject in our field, it is fitting to start with what economists have written about theirs.

**HOME BIAS IN ECONOMICS**

If publication patterns in top economics journals bear some resemblance to the publication patterns of top sociology journals, it is worth exploring some the explanations economists offer for why their top journals have malfunctioned. One major category of explanations involves affiliation-based nepotism, and economists generally refer to this type of bias as home bias. Unfortunately, economists refer to many other types of bias as home bias.

The specific meaning of the term 'home bias' in particular can vary independently across two distinct dimensions. Firstly, home bias can refer to two distinct directions of action. The term applies to active actions, such as investors choosing which countries to invest in (Dlugosch, Horn, and Wang 2023), and also receptive actions, such as an editor showing favoritism to submissions to a journal from academics with the same PhD affiliation. Secondly, home bias can refer to either regionally-oriented or institutionally-oriented actions, such as national leaders with an ethnic ties directing resources to their region of origin (Bommer, Dreher, and Perez-Alvarez 2022) and potentially the AJS favoring affiliates of UChicago. We focus on the definition of home bias as referring to receptive and institutionally-specific actions.

There are two direct analogues to the close association between the AJS and UChicago in economics. The first analogue is The Journal of Political Economy, which, like the AJS, is affiliated with UChicago, and a UChicago affiliates are over-represented in the pages of the journal: a full 14.3 percent of published authors in the JPE are affiliated with UChicago (Heckman and Moktan 2020). The second of these analogues, the Quarterly Journal of Economics, is closely affiliated with Harvard but not as closely as the JPE is with UChicago. A full 33 percent of authors in the QJE are affiliated with Harvard and/or MIT (Heckman and Moktan 2020). This is troubling given that it is now the top journal in the field, defined in this case as the journal with the highest median citation count per article (Card and DellaVigna 2013). Publication count, however, is not meaningful in isolation and requires that we know either the origin or outcomes of publications to evaluate claims of putative bias. In other words, to use raw publication counts to investigate journal-level bias, we require a control either in the form of a pre-hoc quantity to serve as denominator or post-hoc quantity to serve as numerator.

Economists of economics have use article-level citation rates as a proxy for article quality and as a form of post-hoc control for this exact purpose. If there were bias in journals hosted at Harvard, for example, we would expect articles from Harvard affiliates to underperform in the pages of Harvard house journals with underperformance defined imperfectly as low rates of average citation. Economists of economics have found that there does not seem to a bias in favor of publications from Harvard affiliates because, as of 2011, Harvard faculty were the most-cited group of scholars in the top 5 economics journals hosted at Harvard (Ellison 2011). There may even be a countervailing bias against authors from UChicago at the UChicago-affiliated JPE, whose articles are cited more in the JPE than those published by the faculty of other departments of economics (Oswald 2008).

This work, while very valuable, does have a few shortcomings. The first is that most studies of nepotism and publication with a few exceptions in the economics of economics have tended to treat institutional affiliation at time of publication and not PhD origin as the main vector of nepotism. This is likely because, until recently, PhD origin was prohibitively difficult to study and had to be manually transcribed from large corpora of academic CVs. The second shortcoming is that no economists have simultaneously examined the separate but related signals of publication count by institution and mean citation count by institution.

**PRESTIGE BIAS**

With the evidence for nepotism in economics established, we may also ask if there are other sources of bias that skew economic scholarship. Nepotism is relational and directed, but there may be a more general and transferable source of unfair advantage conferred by prestige. Economists of economics have explored the impact of prestige on productivity and career outcomes. Some have defined prestige as network centrality, and they have shown that more well-connected advisors lead to better career outcomes (Rose and Shekhar 2023). Others have defined it as rank of graduate institution (García-Suaza, Otero, and Winkelmann 2020). Graduates of top economics programs find full-time professorial employment at higher rates are more satisfied with their academic than their peers of less prestigious origins (Siegfriend and Stock 1999). Students who graduate from prestigious PhD programs attain better jobs than their peers, but there is an independent effect on professional success of being advised by a highly cited advisor even in a highly ranked department (Hilmer and Hilmer 2007). The effect of an eminent advisor can be enormous. Though the definition of 'leading economists' was left vague, a 2000 paper found that the benefit of a 'leading economist' writing a recommendation letter led to a prestige increase of 60 points on the USNWR rankings for a young economist's first professorial job (Krueger and Wu 2000).

PhD prestige also dramatically influences publication success in economics. A landmark paper from 2001 found 28% of publications in top 30 economics journals were published by authors from 12 US-based universities, all of which were highly ranked (Hodgson and Rothman 2001). A replication of the paper found that this concentration was more pronounced for higher-ranked journals; graduates of the top 5 economics PhD programs publish 45.5% of the articles in the top economics 5 journals (Ailstleitner, Kapeller, and Kronberger 2023, Figure 7). The favor shown to graduates of top departments is rendered more suspicious because the median graduate of certain top departments does not appear to perform as well as those from some stand-out lower-ranked departments, including Carnegie Mellon (Conley and Önder 2014). The general atmosphere of inequality and elitism in economics extends into the present. The most recent work on economics journals has concluded that they are "governed by men affiliated with elite universities in the United States" (Baccini and Re 2025).

If elite economics in general is as nepotistic as economists of economics claim and if a handful of key journals with particularist ties to the most elite economic departments in the world most show specific favoritism to other members of their host institutions, it begs the question: how do the other social sciences fare in comparison? Some evidence suggests that politics is no better that economics (Reingewertz and Lutmar 2017). We ask how sociology fares in comparison to its cousins in the social sciences.

**A NASCENT SOCIOLOGY OF SOCIOLOGY**

Where there is an economics of economics, there is barely a sociology of sociology. There is no dearth of sociological thought on science, and the sociology of science is a very rich subfield, but the most eminent sociologists of science have ironically rarely studied their own fiefdoms. Though Long has published primarily in sociological venues for his entire career, his famous series of ASR papers arguing that the main determinants of professional success are not productivity but the prestige of one's job placement and PhD origin focused exclusively on the natural sciences (Long 1978; Long et al. 1979; Allison and Long 1990). Zuckerman's work on Nobel laureates similarly focused on scholars in the harder sciences (Zuckerman 1972; Zuckerman 1977). Zuckerman and Merton do discuss the humanities and social sciences in passing in Zuckerman and Merton (1971), and Stephen Cole does the same in Cole (1983), but these works are inherently comparative and seek to understand the hard-soft distinction. The sociology of sociology never earned the intensive focus of the sociology of science's commanding stars, and it was never a high-prestige pursuit.

The nascent beginnings of a sociology of sociology may be found in a handful of papers from the 1970s that presented evidence of elitism and nepotism in the sociological establishment. Many suggested that the main source of bias in the sociology publishing were related to editorship bias. Though the ASR has never been as closely linked to one institution in the manner of the AJS, it has historically been yoked closely to a small handful of very elite institutions. Four-fifths of the 1963-1965 ASR editorial board was from sociology's then-top five departments (Shamblin 1970). From 1948 to 1968, PhD alumnae of only 3 institutions, namely UChicago, Harvard, and Columbia, commanded 61.2 percent of the positions on the ASR editorial board (Yoels 1971). During this period of the ASR's history, editors-in-chief from one of three aforementioned departments tended to recruit alumnae of their PhD-granting department to the editorial board at disproportionate rates, and it was not uncommon for a full 33 to 37% of the editorial board to be composed of the editors' PhD peers. A quick survey of the ASR editorial board proves that this is the case: while the editors of the current ASR editorial board are currently professors at the University of Massachusetts-Amherst, the full editorial board features scholars from universities across the United States and even from most major regions of the world, and the rankings of these universities vary dramatically.

While there is some contemporary work on the sociology of sociology, there is very little, and it does not address bias in sociology's top journals. The closest equivalent to The Journal of Economic Literature in sociology is The American Sociologist, but the 2 journals differ greatly in their impact and prestige. The Journal of Economic Literature's impact factor is 10.6 as of 2024 (Clarivate 2025), and it serves as an informal repository for review articles of economics as well as a forum for articles on the state of economics. There is, then, an incentive to publish in the journal, and many of the biggest names in economics have published economics of economics articles in its pages. This is not the case for The American Sociologist, which has an impact factor of 1.1 as of 2024 (Clarivate 2025). While the impact factor is an imperfect metric, articles in the journal are not very visible to sociology as a profession and do not garner much attention even though many of the sociology of sociology articles it publishes are of very high quality.

Much modern research on stratification in sociology has been conducted with a relational, network-oriented focus. It is generally the case in sociology that elite PhD programs overproduce graduate students who then must seek tenure track professorial posts at less prestigious universities (Baldi 1994). Some sociologists have followed the implications of this finding and adopted a framing of prestige defined as "centrality within hiring networks" (Val Burris 2004, 239; Hevenstone 2008). This view is able to explain a very large amount of the variance between departmental rankings, but it relies on outcomes to define prestige. Prestige is at least as much about public perception as objective outcome, and it may be more fitting to use measures that directly capture how sociologists view their peers, such as the US News and World Report rankings, and some work has addressed prestige in the modern sociological establishment. PhD prestige has immense influence on the prestige of an academic's first professorial job (Baldi 1995).

**EVIDENCE OF BIAS IN SOCIOLOGY'S TOP JOURNALS ACROSS TIME**

Sociologists have investigated bias in top sociology journals from the discipline's earliest decades to the 1970s, but they have not done so since. For a sense of how these trends have changed over time, we can compare the overrepresentation of UChicago PhD alum in the pages of the AJS from the journal's earliest days to the present. The AJS was seen as an informal operation, and, perhaps as a consequence of its informality, it was also widely regarded as a dominion of UChicago PhD alumnae (Abbott 1999). The nature of competition suggests that more competitive markets permit less room for bias to muddy results (Hong and Kacperczyk 2010), and the number of sociologists has increased dramatically since the 1900s. We can then assume that bias in sociology's top journals was mechanistically more pronounced in the early half of the 1900s, and we can take the bias of the AJS shown towards UChicago graduates and alumnae of top institutions, which is to say the house bias of the AJS and the prestige bias of the AJS and other top journals, as a baseline for evaluating the bias of sociology's major journals in the present.

There are two sources for the PhD origins of authors in sociology's top journals from before the period that coverage that we study, which is 1970 to the present. The first of these sources on early bias in the AJS covers 1900 – 1920. Table 2 in Abbott (1999, 90) records the PhD origins of contributors to the AJS for the first three full decades of the journal's history. The table is summarized below with authors from Columbia, the second most-represented school, included for completeness.

[Table 1 about here]
[Table 2 about here]

Overall, we see that alumnae from UChicago published between 14.6% and 22% of the articles in the AJS from the 1900s to the 1920s.

The second historical datapoint on bias in sociology's top journals is from 1945 a study of ASR submissions, which found that submissions from elite institutions both private and public were more likely to be accepted than those from less-prestigious institutions (Goodrich 1945).

The author of this piece unfortunately used then-current affiliation instead of PhD origin, which means that it is not directly comparable to Table 1 and Table 2. Regardless, this still provides evidence of prestige bias in sociology's past. The work also showed that universities from the mid-Atlantic submitted 45 articles with 21 articles accepted, and those from the eastern portion of the Midwest submitted 35 with 24 articles accepted. Sociology departments in these two regions occupy the commanding heights of elite programs in sociology. Universities from the American South, a region with a comparative dearth of strong sociology programs, submitted 7 articles with 5 articles accepted, and those from outside the US submitted 7 with 6 articles accepted (Goodrich 1945). We see, then, that authors from regions of higher average prestige submit many times more articles but enjoy a far lower rate of article acceptance than authors from regions of lower average prestige. The sheer difference in submission volume, however, means that authors from high prestige regions have more than three times the total article count in the ASR. While a wider, more diffuse version of home bias that we might term regional bias may also skew results in favor of authors from the Midwest and the northeast, we can read this as additional albeit inconclusive evidence of prestige bias.

      A third datapoint on bias in sociology publishing covers the early 1960s. In an effort objectively to rank the performance of sociology departments, Knudsen and Vaughan (1969) tabulated the count of articles published in the AJS, ASR, and Social Forces by PhD-granting department for recent sociology PhDs who graduated between 1960 and 1964 and for then-current faculty. UChicago PhD alumnae were the most-published in the AJS by far compared to PhD students from other departments with 133% of the AJS publications of the Columbia, the second most-represented PhD origin in the AJS, and UChicago faculty were the most-published in the AJS by a similar margin compared with the second-ranked University of Michigan.

(Knudsen and Vaughan 1969, Tables 1 and 3). Knudsen and Vaughan unfortunately use a points-based system to weight different types of publications across publishing venues. While this system is interesting, it does not allow for the recovery of raw article counts, and it is hence impossible to compare these results to those tabulated by Abbott.

We now have a rough baseline for later comparison. Regarding home bias, we know that in the 1920s, roughly 20% of articles in the AJS were published by PhD alumnae of UChicago. This home bias had diminished significantly in the 1960s, during which UChicago PhD alumnae published 1/3 more articles than alumnae of the second-most published PhD origin. Regarding prestige bias, we know that all journals published far more work from sociologists of prestigious PhD origins, though we do not have a measure of the discrepancy in publishing success between prestigious and less-prestigious PhD origins.

## DATA AND METHODS

### DATA

We are the beneficiaries of access to data that previous scholars in this field did not enjoy. Large bibliometric databases, such as the Web of Science and Scopus, have brought big data to the study of science and transformed the quantitative study of knowledge production.

We draw on a combination of databases with overlapping but distinct bits of information and link them. We specifically use the Microsoft Academic Graph, which contains information on more than 150 million papers and their authors; the ProQuest Dissertations & Theses database, which provides access to the full text of almost every dissertation published in the US and many published abroad since 1970 with associated metadata; and Scopus, an index of journals that provides abstracts for most articles. We focus on PhDs from the United States, and we do so for a few reasons. Sociology in the United States is notably parochial in terms of both disciplinary and national focii. While there are some scholars who have earned doctorates in other fields, they are rare. Similarly, while there are some scholars from outside the US who publish in top American journals, notably a contingent of Dutch scholars, the vast majority of those who publish in American sociology journals earned their doctorates in sociology from American universities and publish almost exclusively on the United States. The content of American sociology tends to revolve around the United States.

We link the ProQuest Dissertations & Theses database to Microsoft Academic Graph using a combination of string matching and vectorized representations of words via word embedding methods. We specifically use GloVe (Pennington, Socher, and Manning 2014). We supplement the MAG records with more detailed journal-level information on articles in the AJS, ASR, and Social Forces from Scopus. This gives us a comprehensive view of all the publications of every scholar who has published in the AJS, ASR, or Social Forces from the 1970s to the present.

We first present a selection of descriptive results. Figure 1 shows the cumulative percentage of articles published by PhD-granting sociology departments arranged in order of article count in top journals. It is an application of the Lorenz curve to the social sciences with articles in top journals swapped for wealth and PhD-granting departments swapped for households or individuals (Lorenz 1905). We see that the inequality of each journal as expressed by the relative monopolization of journal publications by top institutions varies dramatically by journal. The AJS is the least equal journal among sociology's top 3 journals. The ASR is slightly more egalitarian than the AJS, and Social Forces is the most equal of all, but this equality is only relative, for the distribution of articles in these journals to the alumnae of top PhD-granting institutions is far more stratified than even the most inegalitarian countries in the world.

**METHODS**

After linking our databases, we tabulate a series of index-based metrics to control for relative bias inspired by Greenman and Xie (2008). Before formulating these metrics, we must decide how to quantify bias. The central question that emerges from the extant literature in sociology and economics on both prestige bias and home bias in journal publishing is where in the publishing pipeline to look for bias.

[Table 3 about here]

Each stage of the publishing pipeline offers a new metric that may be used to examine bias. We offer a summary of these metrics and their potential significance in Table 3. Some of these metrics are already proxies of article quality. Acceptance rates are a pre-hoc proxy for quality, and citation counts are a post-hoc proxy for quality. More specifically, acceptance rates loosely map to quality as defined by the editors of a given journal, and citation rates are a very imperfect measure of quality as evaluated by a field or subfield. We may then combine metrics from various stages of the pipeline to examine how papers move from submission to publication in more detail.

[Table 4 about here]

Table 4 provides examples of some of these metrics. We would ideally have access to the relationship between editorial acceptance rate and citation by PhD origin, but this requires access to very sensitive data. We focus on the relationship between publication decisions and post-publication citation count because it relies on easily accessible information on articles and offers a reliable way to scan for bias in editorial decisions.

After deciding upon which stage of the publication process to direct our attention, we may now tabulate indices of publication and citation. We must first, however, contend with the primary difficulty of comparing citation counts across any unit of analysis, which is that the distribution of citation counts may be approximated by the power law distribution because citation counts are distinguished by extreme outliers (Clauset et al. 2009; Peterson et al. 2010; Brzezinksi 2015). While the arithmetic mean is serviceable for normal distributions, the arithmetic mean of a power law yields a very biased estimate skewed towards higher values. These problems only grow more pronounced when analyzing subgroups of a power law distribution, such as the average citation per PhD-granting institution of articles in a given journal. A possible solution may be found in the geometric mean. For citation counts specifically, the geometric mean is more reliable than the arithmetic mean or percentile for comparing citation counts between nations (Thelwall 2016). We adapt the spirit of this approach to compare citation counts between articles written by alumnae of different PhD-granting institutions. The geometric mean has a very long history of usage in the social sciences (Galton 1879; Goodman 2017). While the geometric mean is more mathematically sound in theory than the median for comparing citation counts, the geometric mean and the median behave very similarly in practice. Other scholars (Card and DellaVigna 2013) have used medians to compare citation counts. We perform the analysis with both methods, and our analyses are robust to this specification (see Supplemental Information).

The geometric mean is found by taking the arithmetic mean of a logged sequence of numbers and then exponentiating the result:

$$\text{Geometric Mean} = \exp\left(\frac{1}{n}\sum_{i=1}^{n} \ln(x_i)\right)$$

A second major difficulty that arises when working with citation counts is that citation counts vary between journals and across time. To control for this, we propose an index that we term the citation multiple. The metric answers a simple question: How do we establish whether articles from PhD alumnae of a certain school over- or under-perform relative to articles from alumnae of other institutions? We control for periodicity and journal effects by taking the geometric mean of the citations earned for all articles in a given journal in a given decade. We then match each article to its journal- and decade-specific mean and divide the citations said article has earned by the journal- and decade-specific geometric mean of citation. This gives us a measure of relative citation count. We then take the geometric mean of citation multiple by institution. This yields a robust measure of how the progeny of a given institution compare to the those of other institutions. An institution-level citation multiple of 1 indicates that PhDs from said institution publish articles that perform as well as the average articles in top sociology journals. An institution-level citation multiple of 2 would suggest that PhDs from said institution publish articles that perform twice as well as what would be expected.

Over- or under-citation, however, is only one component of bias in sociology publishing. If publishing in the sociology's top journals is the central currency by which young academics either advance or falter in the academy, then the binary outcome of publishing in sociology's top journals or not is what matters. It is possible to create an index in the vein of the citation multiple to investigate over- or under-publication by PhD-granting institution. Because publication counts do not follow a power law distribution, calculating publication overrepresentation is much more straightforward and does not require geometric means. The simple arithmetic mean is adequate for comparing publication counts between institutions.

We now have indices for over- or under-citation and over- or under-publication. While we can use these indices in isolation, we can combine them to arrive at a single, readily interpretable metric that summarizes both aspects at once. By dividing the publication multiple by the citation multiple, we can plot over-publication and under-citation simultaneously. Given that we expect PhD alum from prestigious universities to be both over-published and under-cited in sociology's top journals, this metric is ideal for our purposes, and we term it the *citation-adjusted publication premium*. We use this metric to explore two specific manifestations of elitism in sociology: a general bias in favor of prestigious PhDs from prestigious universities across all journals that we term *prestige bias* and a more specific bias in house journals towards PhDs from affiliate departments that, in an extension of the econometrics literature, we term *house bias*.

**RESULTS**

We begin by presenting our results for prestige bias, the most general form of bias. We first segment the prestige of PhD origins as quantified by the US News and World Report into three band: Top 1-15, Top 15-50, and Top 50-100. Economists of economists have used similar approaches to group schools by prestige (Siegfried and Stock 1999).
The relative positions of sociology departments have changed dramatically over the past five decades. To account for this, we have collated a list of history USNWR rankings and match every PhD graduate to the prestige of his or her department at year of graduation[1].

---

[1] For years with missing data, we interpolate USNWR rankings by taking the average of years on either side of the rankings lacuna.

[Figure 2 here]

Figure 2 shows the change in prestige bias over time for the AJS, the ASR, and Social Forces. It is immediately evident that the citation-adjusted publication premiums for all three prestige bands are very different remain separate except for only one of 15 possible journal-decade combinations: in the AJS in the 1970s, the citation-adjusted publication premium of Top 10-50 departments is lower than that of Top 50-100 departments for the only time in our data. Apart from this lone exception, Top 10 departments enjoy a strong and pronounced advantage citation-adjusted publication premium that peaks in the 1990s and 2000s and declines in the 2010s. This decline in the 2010s is most pronounced in the AJS, but the ultimate result of this decline yields a citation-adjusted publication premium of 1.34, which is composed of a publication multiple of 2.00 and a citation multiple of 1.49. In other words, top 10 schools published papers twice as often as would be expected, but their work was cited only 1.5 times more than the work of all other schools. The bias towards top 50+ institutions, however, remains almost unchanged in the AJS into the 2010s. PhD alumnae from top 50+ institutions publish 0.38 papers for every 1 paper published on average, but their papers are cited 1.91 times more often than average. This publication multiple of 0.38 and citation multiple of 1.91 yield a citation-adjusted publication premium of 0.20. It is of interest that the ASR maintains, on average, the most intense prestige bias in favor of top 10 institutions, who enjoy a citation-adjusted publication premium of 2.01 in the ASR in the 2010s, and against top 50+ institutions, who suffer from a rock-bottom citation-adjusted publication premium of 0.18 in the ASR in the 2010s. The least prestige-biased journal is Social Forces.

A decomposition of the citation multiple and publication multiple plots used to create the plot of the citation-adjusted publication premium may be found in the Appendices (Figures S1 – S2). We additionally provide a diachronic view of prestige bias where we present prestige as a continuous variable instead of the discretized variant we use in the analyses above (Figures S3 – S5).

**HOUSE BIAS**

We have seen that, in the aggregate, the citation-adjusted publication premium is inversely related with departmental prestige. To examine house bias, we shift our focus slightly from prestige bias among departments of all prestige levels to house bias among a subset of the most elite sociology departments. We do so because prestige bias is distributed unevenly throughout all schools, albeit very unevenly, but house bias is a nepostic backdoor that offers a single department a backdoor route to publish in a single journal and necessarily offers a much weaker signal.

We shift our focus from the ranking of all schools based on prestige to the top 10 schools as defined not by prestige but instead by publication count in sociology's top 3 journals. The main dependent variable of interest for prestige bias is departmental prestige, but the main dependent variable interest for house bias is publishing quantity. If we assume that the schools with house bias backdoor connections publish the most in the journals to which they are connected, then we should compare schools directly on the basis of publication count.

[Figure 3 here]

Where prestige bias fell in the AJS in the 2010s, we see the opposite trend for house bias. Figure 3 shows that the favoritism given to UChicago PhD alumnae in the pages of the AJS as measured by the citation-adjusted publication premium declined from the 1970s to the 1990s, but it then increased dramatically and reached a peak of 4.45 in 2010. This citation-adjusted publication premium is composed of a publication multiple of 2.62 and a citation multiple of 0.22. This means, then, that UChicago PhD alumnae published 2.62 times more papers than PhD alumnae from the other top 9 institutions but that the papers of UChicago PhD alumnae were cited only 22% as often. By contrast, the average citation-adjusted publication premium for the other nine departments in the top ten most-published departments remained roughly constant. A handful of schools are outliers with citation-adjusted publication premiums above that of UChicago in some decades, but, in the 2010s, UChicago rockets to first place.

[Figure 4 here]

In contrast to the AJS, house bias in Social Forces as measured by the citation-adjusted publication premium (Figure 4) declined dramatically from a high of almost 3.5 in the 1970s to 1.5 in the 2010s. UNC-CH alumnae have not been ranked first on this metric since the 1980s. In relative terms, UNC-CH has been the third school as ranked by citation-adjusted publication premium from the 1980s to the 2010s.

Overall, we see differing pictures of the progression and relative rates of bias in each journal. In the AJS, prestige bias declines while house bias increases dramatically. In Social Forces, both types of bias decline over time. The affiliation of the ASR with the American

Sociological Association instead of any particular institution means that it cannot support house bias, but it is the most sharply prestige-biased of any of sociology's top 3 journals.

The primary limitation of our work thus far is that graduates of different departments publish proportionally more often in different subfields and that these subfields have sharply differing rates of average citation. This limitation has two parts, neither of which have yet been empirically verified.

Regarding average citations, it is a common contention among sociologists that different genres of sociology are cited at varying rates, and this is certainly true. Generally speaking, papers that with stereotypically 'softer' methodological or topical foci are cited less, and papers that read more like those from the 'hard' sciences are cited more. This is perhaps because scholars in 'harder' fields generally publish shorter papers at higher frequencies than do scholars in 'softer' fields (Cook and Xie 2023). While empirically proving the following hypothesis is beyond the scope of this paper, sociological work with high citation counts likely attracts cross-disciplinary citations from harder fields. The differential in paper volume between the hard sciences and soft sciences means that the easiest way to garner more citations is to write a paper that scholars in a 'hard' science will cite. A quick glance at highly cited articles in sociology suggests this is the case. The most cited article ever published in the American Journal of Sociology is *The Strength of Weak Ties* (Granovetter 1973), which is cited extensively for work on networks in all fields. Granovetter (1973) has almost 75,000 citations, and the majority of these citations are from fields outside of sociology. For additional evidence of this, see the Appendices (Figure S6). Regarding differences in genre by department, it is widely known, for example, that sociologists from UChicago and Berkeley are known for producing qualitative work on atypical topics (Abbott 1999, Burawoy 2021). Many sociologists have intuit that this work garners far fewer citations than demography-adjacent sociology, economic sociology, or computational sociology, and this appears to be the case (Figure S7).

Given these two points regarding citation by topic and differences in topic by department, one might think that the house bias shown to UChicago in the AJS can be explained by UChicago PhDs focusing on under-cited genres of sociology. To address this, we control for topic by creating a citation multiple that is topic-specific as well as decade- and journal-specific. While the direction of every result discussed above is robust to this additional specification, controlling for topic does change the magnitude of some of the above results and results in the reduction of most types of bias globally (Figures S8-S10).

The main thrust of our topic modelling approach is not to accurately model the Byzantine subdisciplinary fissures of academic sociology, an exceptionally rich and high-dimensional academic discipline replete with complicated patterns of word usage within and between subdisciplines. Corpora with characteristics like this are precisely where topic modeling approaches tend to break down the most quickly, and addressing the issues inherent in working with such texts is an active area of research in natural language processing in both applied and foundational contexts (Eshima, Imai, and Sasaki 2024; Pham et al. 2024). Our intent is instead to use the output from topic modeling as an additional control in the index-based inequality analyses to add additional robusticity.

With the above caveat established, there are two main limitations to our topic modeling approach. The first it is that the topics in our topic model do not capture many of the aspects of sociological inquiry that are of the most interest to sociologists. There is no topic that easily summarizes the offbeat sociology for which UChicago PhD alumnae are famous, and there is similarly no single topic in our model for more mainstream sociological approaches to the quantitative study of inequality. Future work could explore using both supervised and unsupervised classification-based approaches to develop a finer-grained model that more accurately reflects the collective and currently implicit understanding of sociology's many genres. The second limitation of using topic models in this manner is that it may not be substantively useful because it may potentially obscure the politics of taste and topic choice among elite sociologists. Any publication in a top sociology journal is professionally significant, and that graduates of top-tier sociology programs publish preferentially in lowly-cited topics may represent an important signal. Controlling for topic too strongly would then diminish this signal, which is why we have opted for a coarser approach to topic modeling. It may be that graduates of prestigious sociology departments game publication in top journals by knowingly or unknowingly working on lowly-cited topics that are nonetheless still of boutique importance or seen as representing distinguished, 'tasteful' sociology to sociologists (Bourdieu 2002). More broadly, why do the top PhDs of sociology work on these topics? More work is needed to understand what drives sociologists who were trained at the most eminent departments to study what they study.

  As a final check on the health of sociology's top journals, we may briefly ask how many years into their academic careers sociologists publish in these journals. There is evidence that tenure-track faculty publish more pre-tenure than post-tenure and that the rate of publication increases up until tenure evaluations across all of the sciences (Tripodi et al. 2025). Given the special place of top 3 publications in sociology hiring, it is worth asking if the rate of publications in sociology's top 3 journals map to tenure timelines. We may again turn to economics for an analogy with our own field. The brutal competition for space in the top five journals has rendered publishing in them a young scholar's game. Established scholars are more likely to use pre-print venues to circulate their work than brush elbows with plucky graduate students or untenured faculty in the narrow bottlenecks to top journals (Ellison 2011). We find that this is also true for sociology. The age structure of sociology journals is similarly skewed towards the young; the mean academic age of publication in sociology's top 3 journals is, depending on the journal, 7 to 9 years after acquiring a PhD (Figure 5). Viewed thusly, sociology's top journals, then, are not only the discipline's pre-eminent fora for sharing knowledge; they also collectively serve as a "screening device" (Heckman and Moktan 2020, 462) for young scholars. This does not mean that high-quality and/or foundational papers do not end up in sociology's top journals, for it is readily apparent that they do. It is instead to suggest that the processes that ultimately enable useful knowledge to accumulate in journals may be fruitfully optimized.

**DISCUSSION**

Publishing in sociology is an elitist venture. Prestige and nepotism both tilt the scales of landing a paper in one of sociology's top 3 journals. Many sociologists have wondered if there is home bias between the two house journals in sociology, namely the UChicago-affiliated AJS and the UNC-CH-affiliated Social Forces, and the institutions that host them. This is unfortunately the case. The works of UChicago PhD alumnae are over-published but under-cited in the AJS, and this favoritism has generally increased over time. While the works of UNC-CH PhD alumnae PhD were once over-published but under-cited in Social Forces, this bias has diminished to essentially zero as of the 2010s. More broadly, the works of scholars with PhDs from prestigious institutions are similarly over-published and under-cited in the pages of all of sociology's top 3 journals. This prestige bias has diminished over time in all journals.

These results have both methodological and substantive implications for the study of science in sociology and beyond. The first is that the methodological toolkit for tracking affiliation- or prestige-based publication bias outlined in this paper can be readily deployed to other disciplines. A key strength of the index-based approach outlined in this article is that does not rely on proprietary data that must be provided by individual journals. Obtaining data that is not publicly available, such as article acceptance rates, may require prohibitively time-intensive negotiation with the staff of different journals, and gaining more sensitive types of data, such as fuller metadata on the authors that submit to prestigious journals or any data on the peer review process, is even harder and risks the violation of academic privacy norms. The index-based methodological approached behind the publication and citation multiples is powered by metrics that may be gleamed from any bibliometric database, namely publication count and citation count. While this work groups by authors by PhD origin, which does rely on a proprietary database, it is possible to use the citation and publication multiples at other levels of analysis to track other types of bias, such as biases favoring certain institutional affiliations for professors or even gender- or ethnicity-based bias. Finally, this paper's use of topic modeling output as a covariate for analyzing inequality represents an extension of extant topic modeling approaches and inform future applications of text data to the study of inequality within or without science.

The substantive implications of this work may be grouped into practical and theoretical concerns. The practical concern is that the sociological establishment has failed to live up to some of its most dearly cherished ideals. Sociology is elitist. Many sociologists would likely balk at the implication that sociology is stratified so intensely, but that does not change that it is so. Bias towards elites is not inherently negative. The competition for sociological talent and the wide array of sociological genres means that we cannot realistically expect a world where the prestige of one's PhD has zero influence on publishing outcomes. Sociologists may ask themselves what degree of elitism is acceptable or even beneficial for sociology. More broadly, what are the core values of sociology? Is there any consensus on what a sociologist should believe or do? Sociology is a remarkably pluralistic and low-consensus field. From an organizational perspective, the main virtue of this pluralism is that sociology offers an extraordinarily diverse array of paths to professional fulfillment and niches for sociologists to fill, but its main vice is that there is little to no explicit consensus in sociology on the boundaries of the sociological enterprise or what constitutes good or bad sociology. Because of this confusion over core goals and evaluatory criteria, allocating the spoils of professional success in sociology can be frustratingly difficult. To ensure that new generations of sociologists continue to enter sociology, it may be useful for sociologists to seriously consider how to explicate the evaluative standards of the field, even if only partially.

    Journal publishing in sociology may be flawed, but this problem is not at all endemic nor specific to sociology. This article carries no intention of singling out sociology, for the issues in sociology's top journals outlined in the body of this article and other issues besides may be found in many of the journals that collectively compose the main knowledge preservation mechanism of science. To name a few salient ways in which scientific journals may not be scientific as scientists would like them to be: peer review can be cartelized and gamed, psychology and biology are riven by existentially-threatening replication crises, and fraud is the constant companion and shadow of modern science (Ritchie 2020). It is little wonder that some fields have moved away from the journal system entirely. Preprint cultures powered by arXiv and similar preprint-hosting websites have emerged in many fields, and computer scientists have made advancement in their field contingent on publishing papers in prestigious conference.

Indeed, computer science is the world's first conference-dominant discipline (Cook and Xie 2023), but none of these solutions to remediate the potentially error-prone review processes of journals really work all that well. The problem is deeper than most would like to admit. It is, in short: once knowledge is produced, how can it be transmitted, preserved, and disseminated to ensure that knowledge accumulates as efficiently and as effectively as possible? Scientists have not yet solved this wicked problem. It would bode well for the future of science if they did.

# TABLES

| Decade | Total authors | UChicago PhD count | Columbia PhD count |
|---|---|---|---|
| 1900 | 131 | 29 | 7 |
| 1910 | 164 | 24 | 18 |
| 1920 | 136 | 30 | 30 |

Table 1: PhD origins of AJS contributors by decade from 1900 to 1920 as count

| Decade | Total authors | UChicago PhD count | Columbia PhD count |
|---|---|---|---|
| 1900 | 131 | 22.1% | 5.3% |
| 1910 | 164 | 14.6% | 10.98% |
| 1920 | 136 | 22.05% | 22.05% |

Table 2: PhD origins of AJS contributors by decade from 1900 to 1920 as %

| **Stage of pipeline** | 1: Pre-submission covariates | 2: Decision | 3: Publication | 4: Post-publication citation count |
|---|---|---|---|---|
| **Metric of interest** | Many, but one possible metric is: Total size of PhD cohort | Editorial Decision rates | Publication count | Citation count |
| **Import** | Pre-submission covariates may serve as additional controls for other metrics | A proxy for how the editorial board and reviewers evaluate the pre-publication quality of a paper or set of papers | Because any publication in a top journal is valuable academic currency in and of itself, raw publication count can be a valuable signal of either editorial bias or PhD quality | A proxy for how other scientists view the quality of a paper or set of papers |

Table 3: An overview of the publication pipeline and metrics that may be used for evaluating prestige bias and home bias

| Combined Metric | Source Metrics | Import |
| --- | --- | --- |
| Acceptance rate by PhD origin in a given journal | 1: Pre-submission covariates [size of cohort] + 2: Decision | A confounded measure of both PhD student quality and potentially prestige or house bias; requires sensitive data |
| % of PhD alumnae from a given PhD origin with a publication in a given journal | 1: Pre-submission covariates [size of cohort] + 3: Publication | A confounded measure of both PhD student quality and potentially prestige or house bias |
| Relationship between publication count and citation count for given PhD origin in a given journal | 3: Publication + 4: Post-publication citation count | Combines publication count, a strong professional signal, with citation count, a signal of quality with some professional importance, to examine editorial judgement post-hoc |
| Relationship between editorial acceptance rate and citation by PhD origin | 1: Pre-submission covariates [size of cohort] + 2: Decision + 4: Post-publication citation count | Combines pre-hoc and post-hoc judgements of quality; would be a very robust measure of editorial bias but requires sensitive data |

Table 4: Examples of combined metrics for evaluating prestige bias and home bias in the publication process

# FIGURES

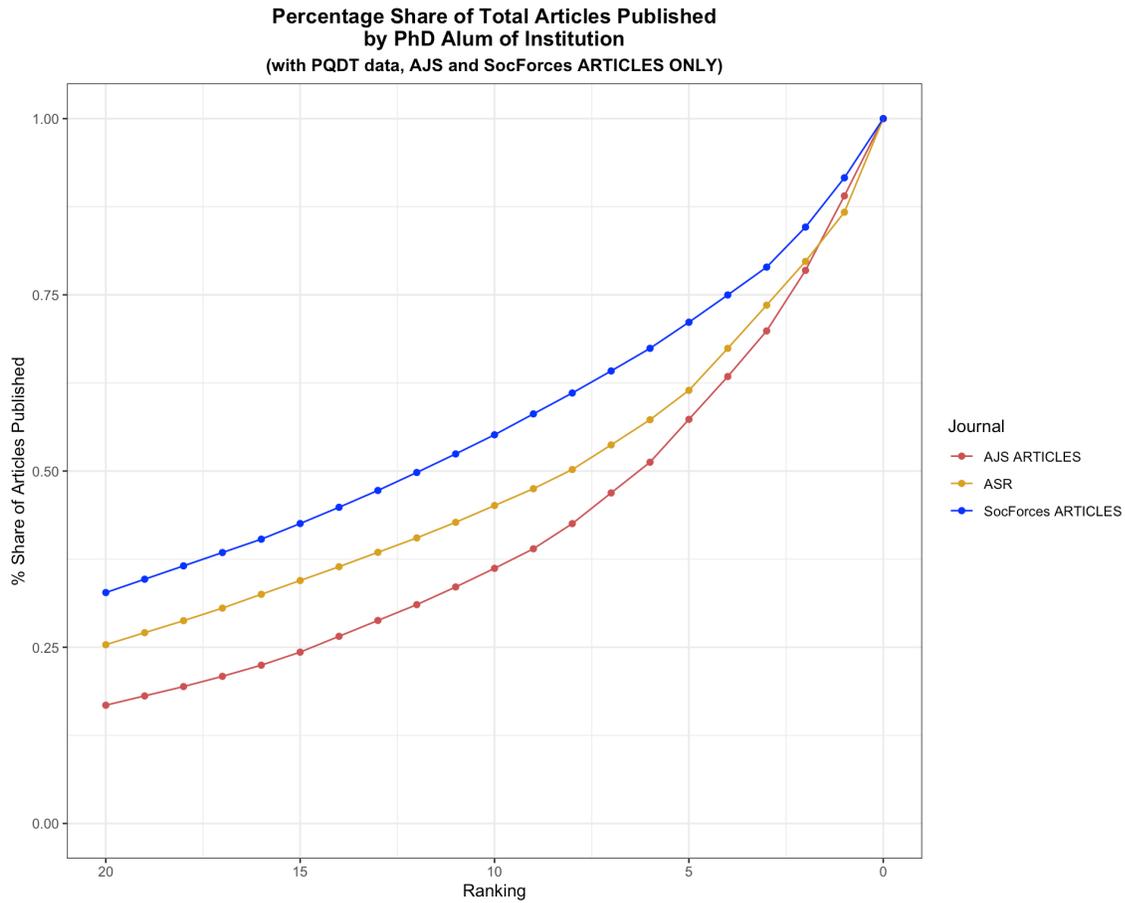

*Figure 1: Lorenz curves of publication count by journal*
Prestigious PhD institutions publish an outsize proportion of the total papers in sociology's top 3 journals. The top 20 institutions by publication in each of the 3 journals author roughly 75 percent of the papers published in each journal. These patterns vary slightly by journal. The AJS is the least equal, the ASR is slightly more equal, and Social Forces is slightly more equal still.

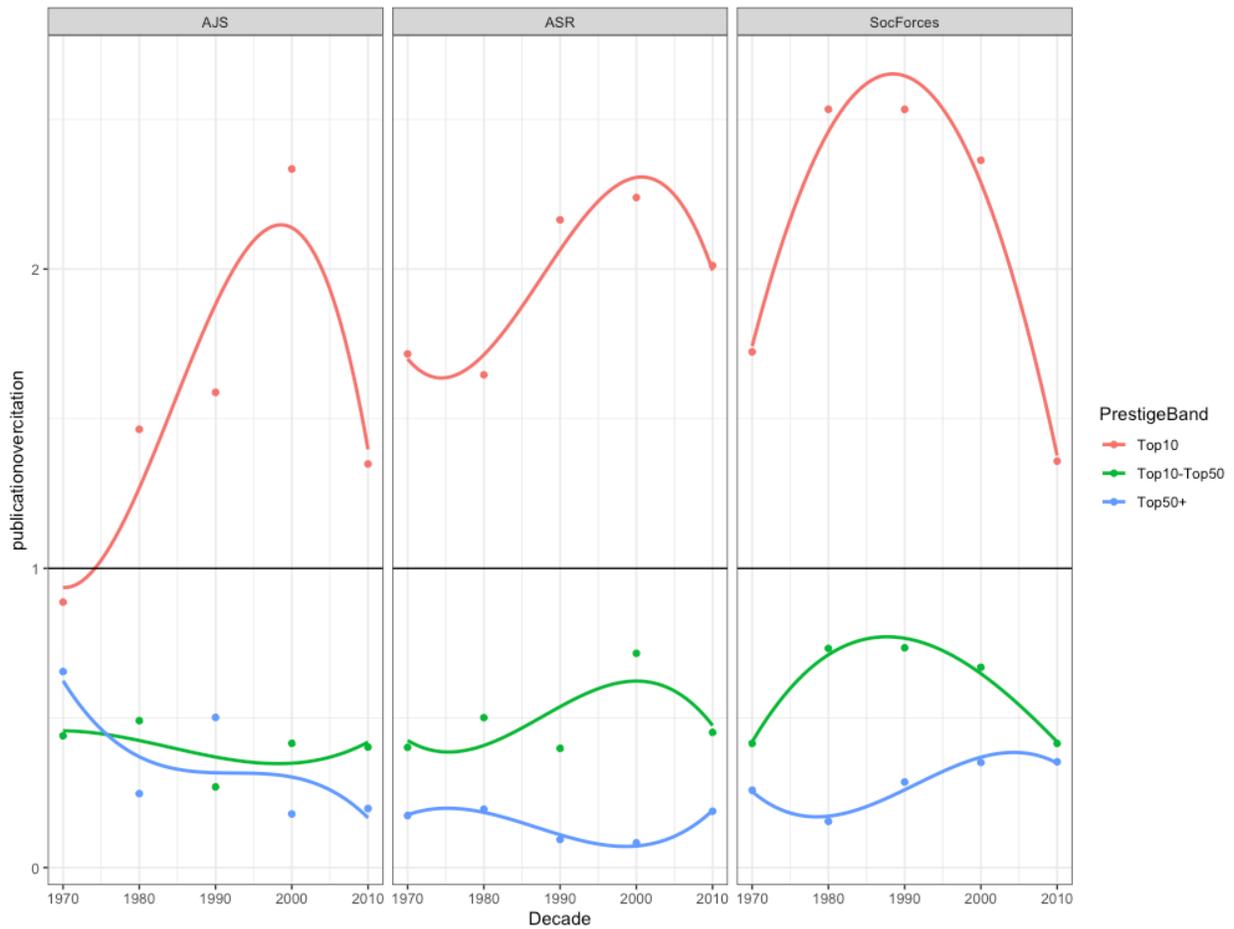

*Figure 2: Prestige bias in sociology's top journals*

We segment PhD origins into 3 buckets: top 10, top 11-50, and top 51 and beyond. With our original metric, the citation-adjusted publication premium, we find that papers from authors with prestigious origins are over-published but undercited in all of sociology's top 3 journals. The bias towards authors from less-prestigious PhDs is mostly constant over time, and the premium accorded to top PhDs improves over time but remains pronounced. The ASR is the most prestige-biased journal and discriminates the most against authors from less-prestigious PhDs.

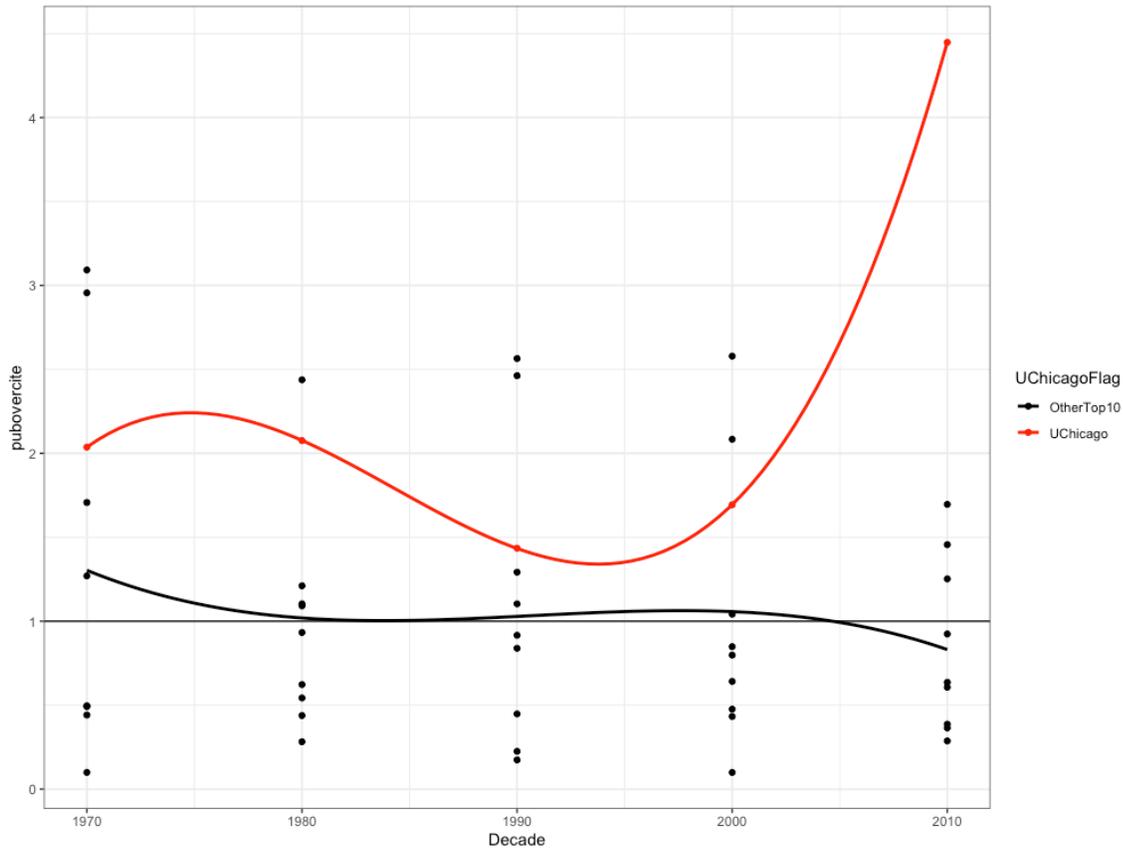

*Figure 3: House bias in the AJS*

We focus on the American Journal of Sociology in isolation. To uncover favoritism that the editors of the AJS may show to UChicago PhD alum, we compare sociology's top 10 programs by publication count in the top 3 journals to one another. We see that AJS alumnae are over-published but under-cited in the AJS far more than the alumnae of any other top sociology program. We also see that this favoritism increases greatly after the year 2000 into the 2010s.

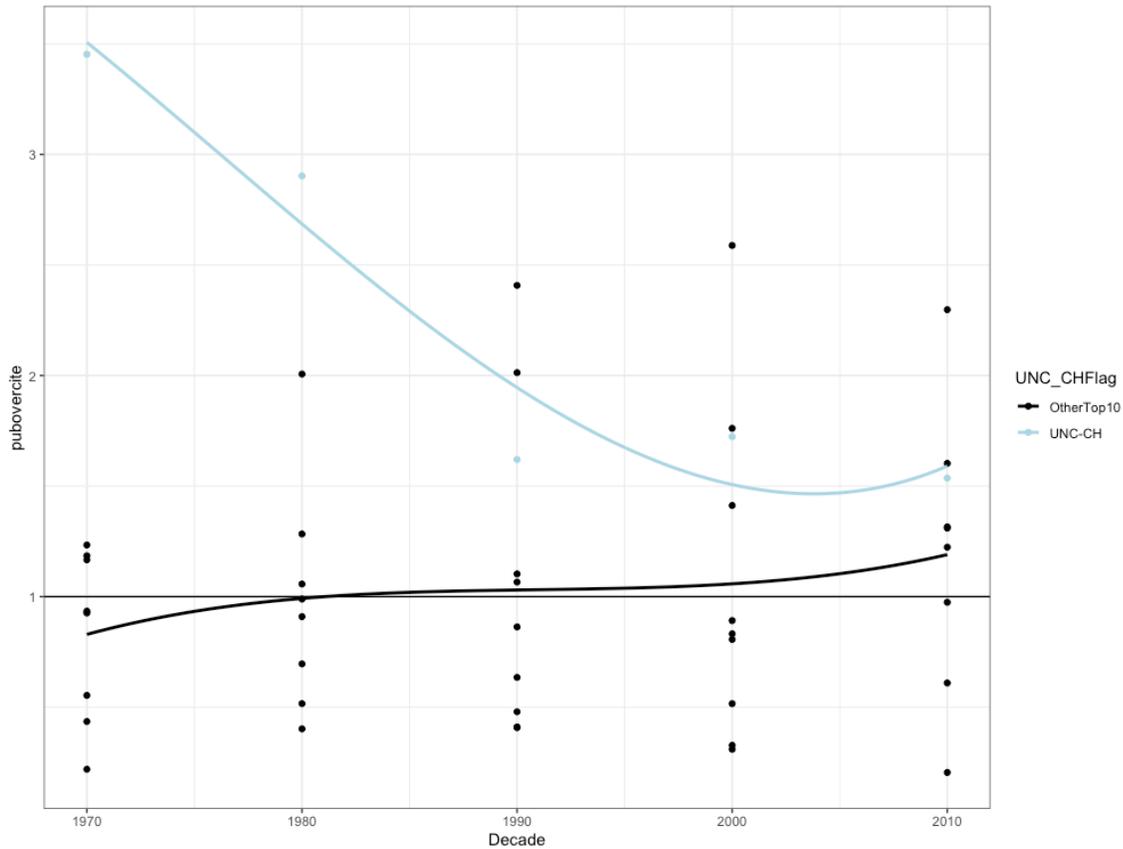

*Figure 4: House bias in Social Forces*

We focus on Social Forces, the other major house journal in sociology, in isolation. As with our analysis of the AJS and UChicago, we compare sociology's top 10 programs by publication count and search for any favoritism that may have been shown to UNC-CH alumnae. While alum of UNC-CH may have been over-published but under-cited in the past, any potential favoritism has greatly decreased over time.

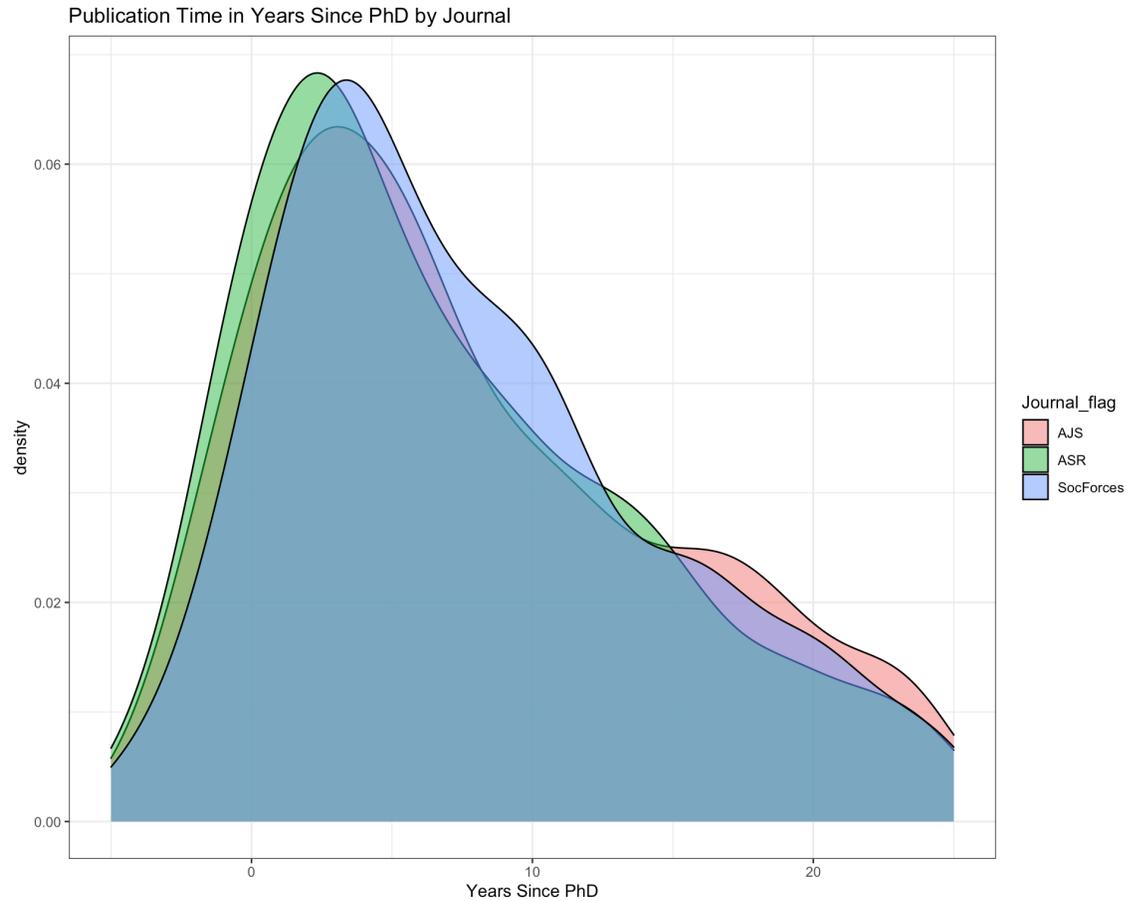

*Figure 5: Age Structure of Publishing in Elite Sociology Journals*
Sociologists publish in sociology's most elite journals, on average, within 10 years of obtaining their PhDs. This suggests that the main purpose of publishing in elite journals is not to advance sociological knowledge. Instead, as economists have found for economics, the main purpose of publishing in an elite sociology journal is likely to secure professional advancement.

# APPENDICES

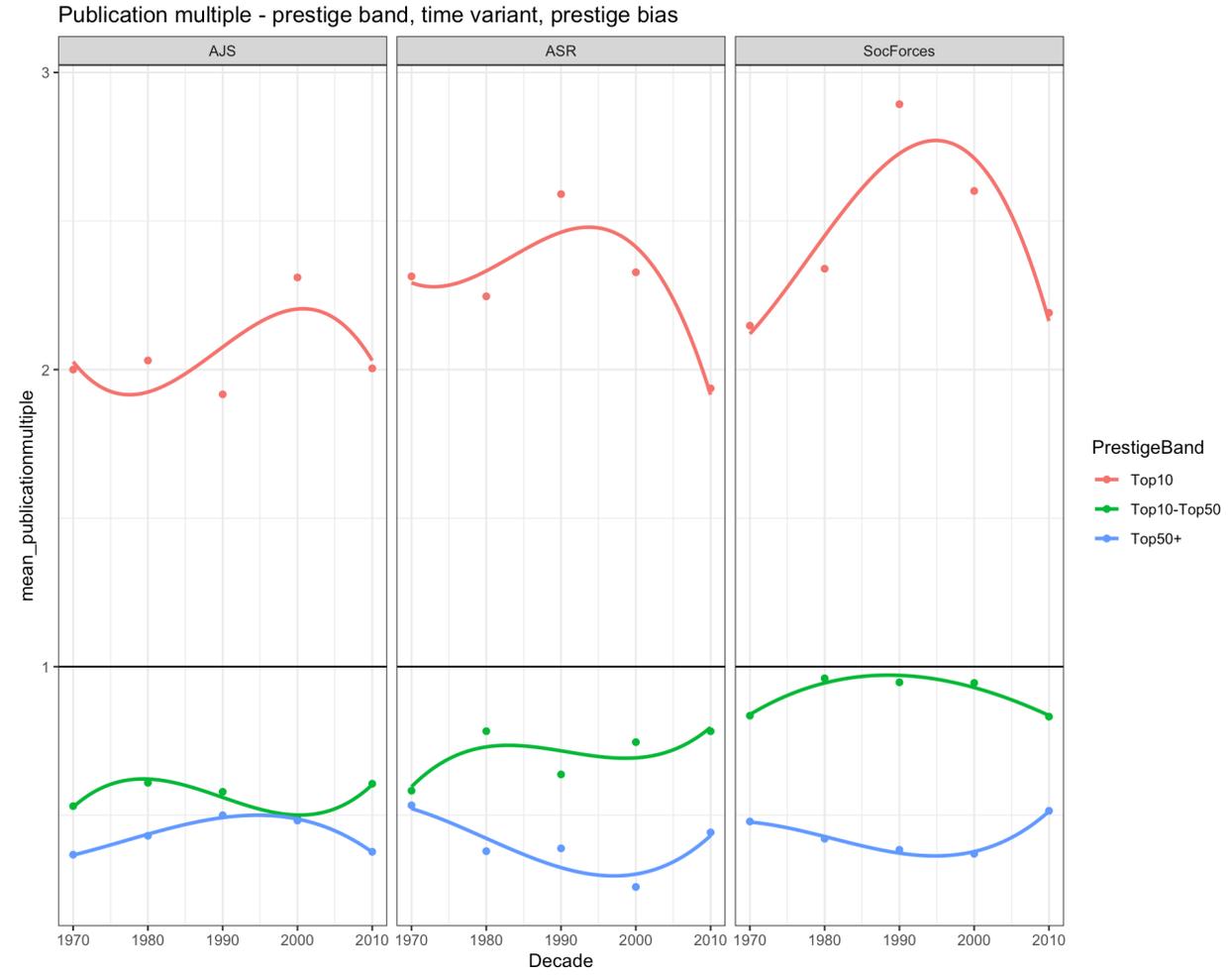

Figure S1: Publication multiple in isolation for prestige bias

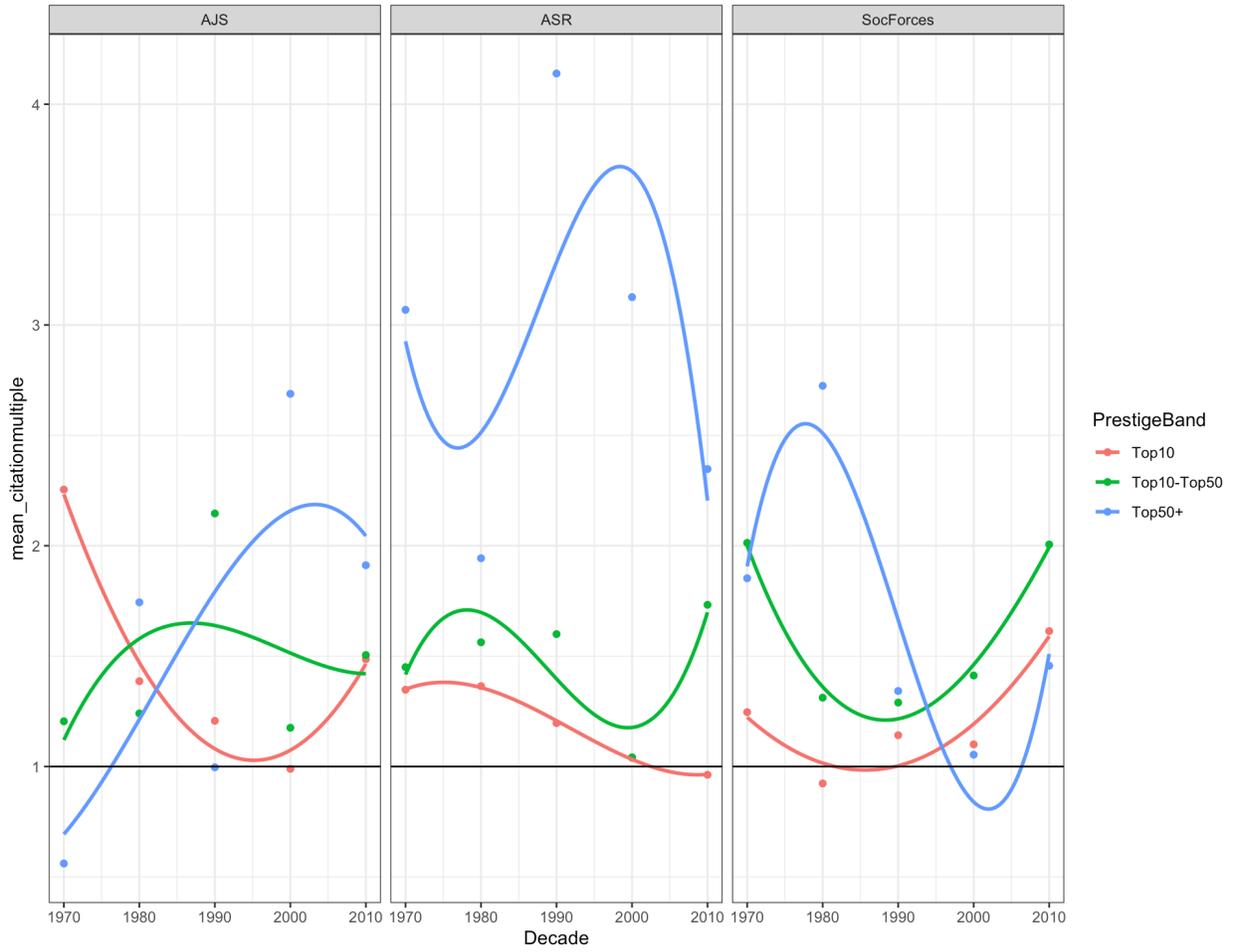

Figure S2: Citation multiple in isolation for prestige bias

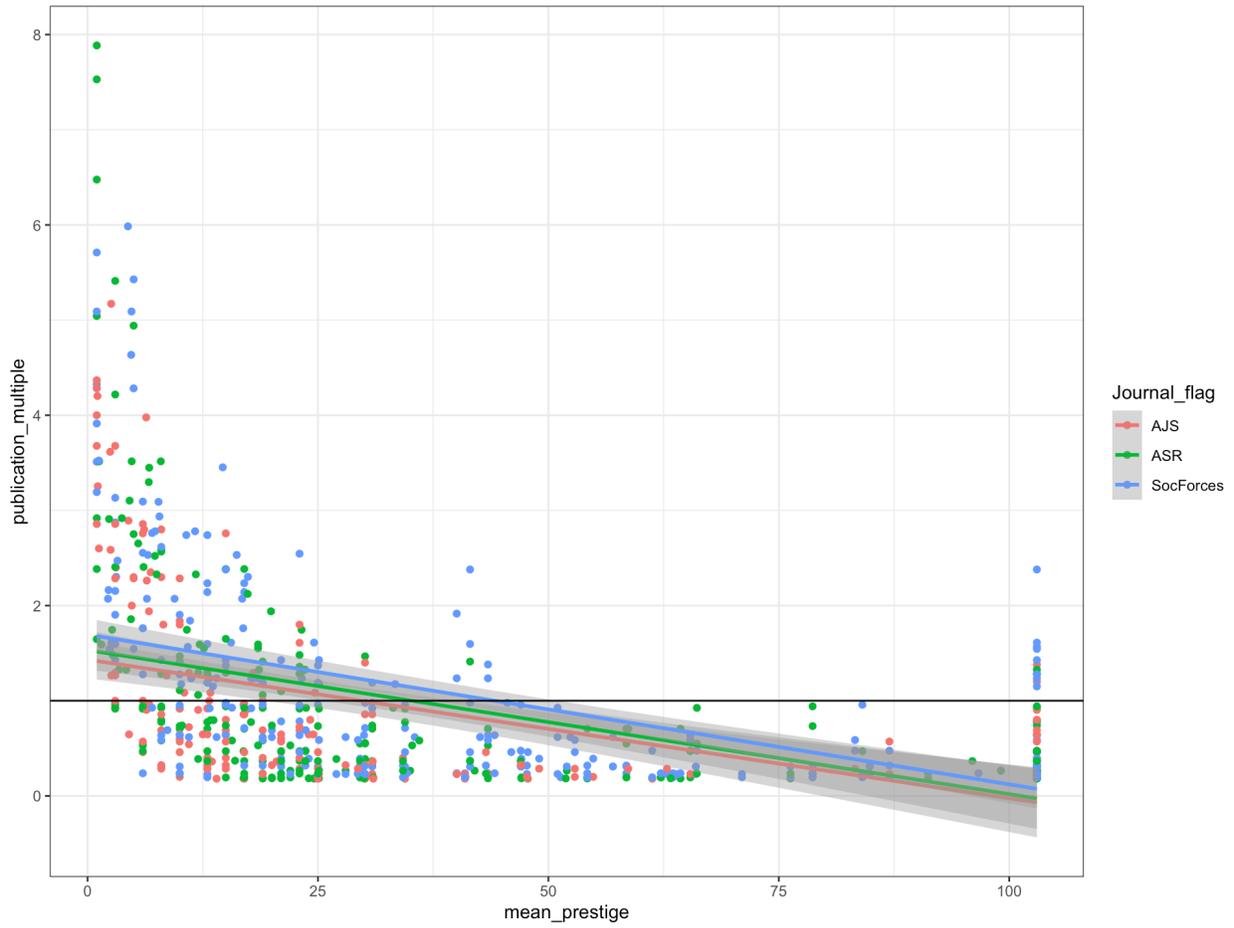

Figure S3: Publication multiple for prestige bias; time invariant

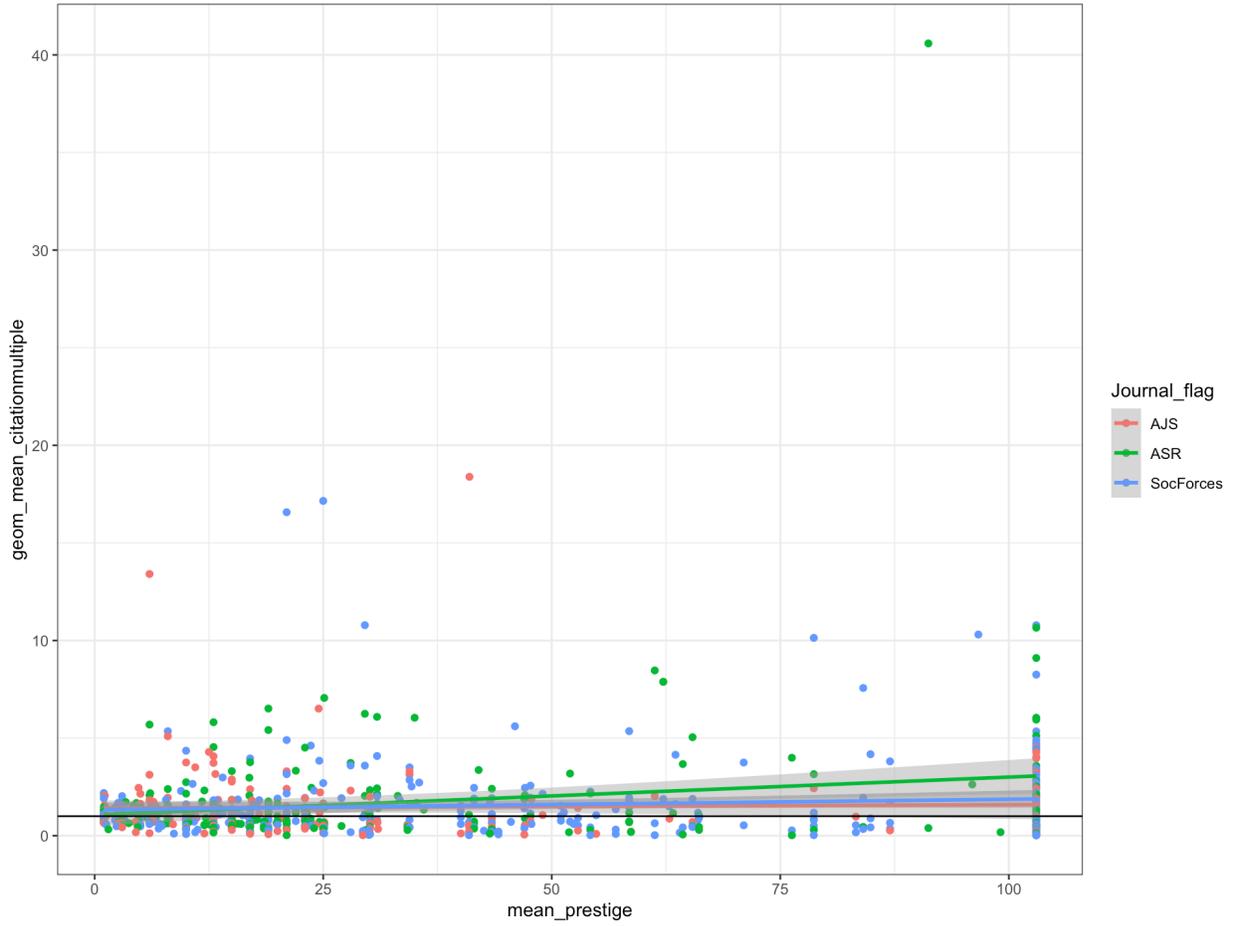

Figure S4: Citation multiple for prestige bias; time invariant

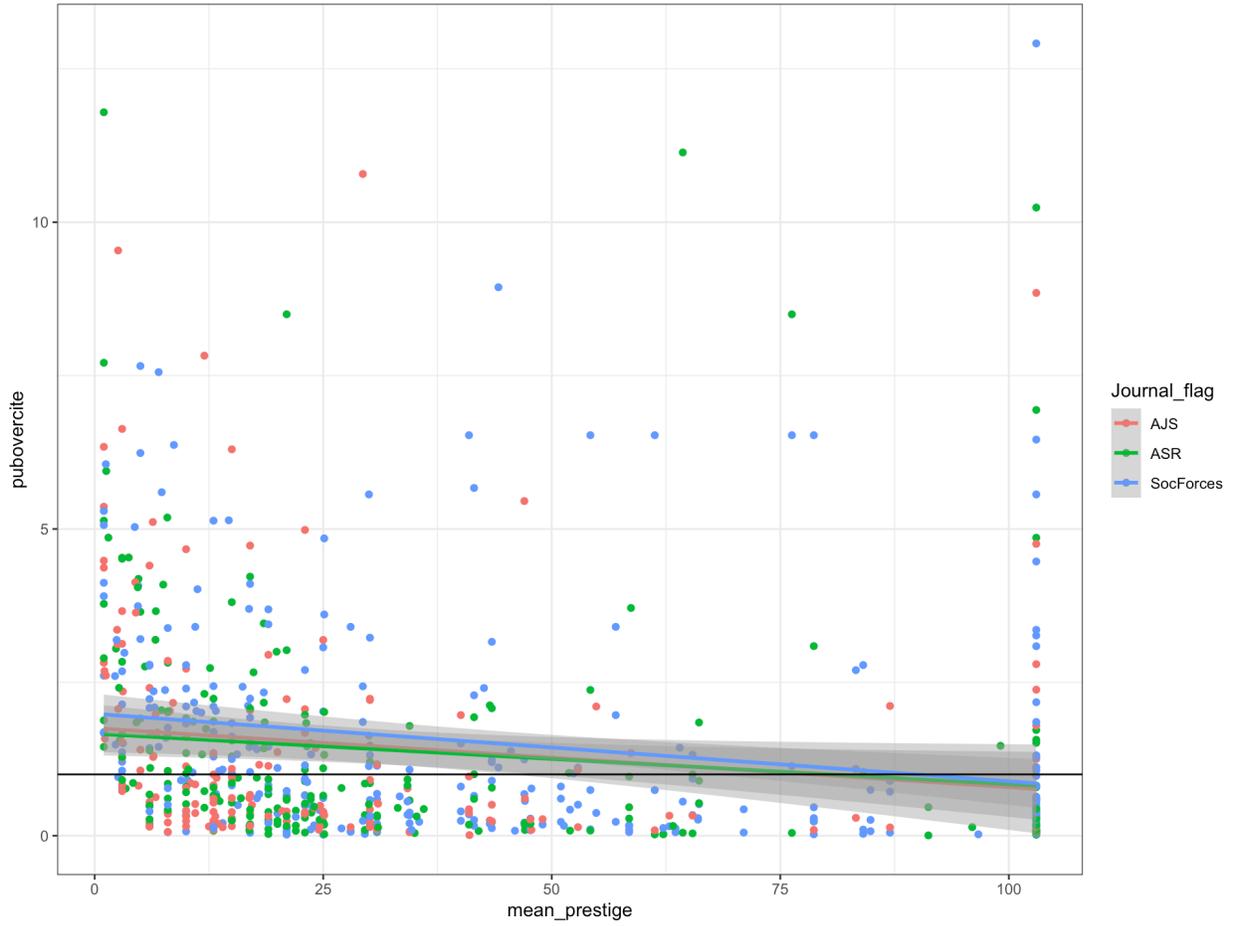

Figure S5: Citation-adjusted publication premium for prestige bias; time invariant

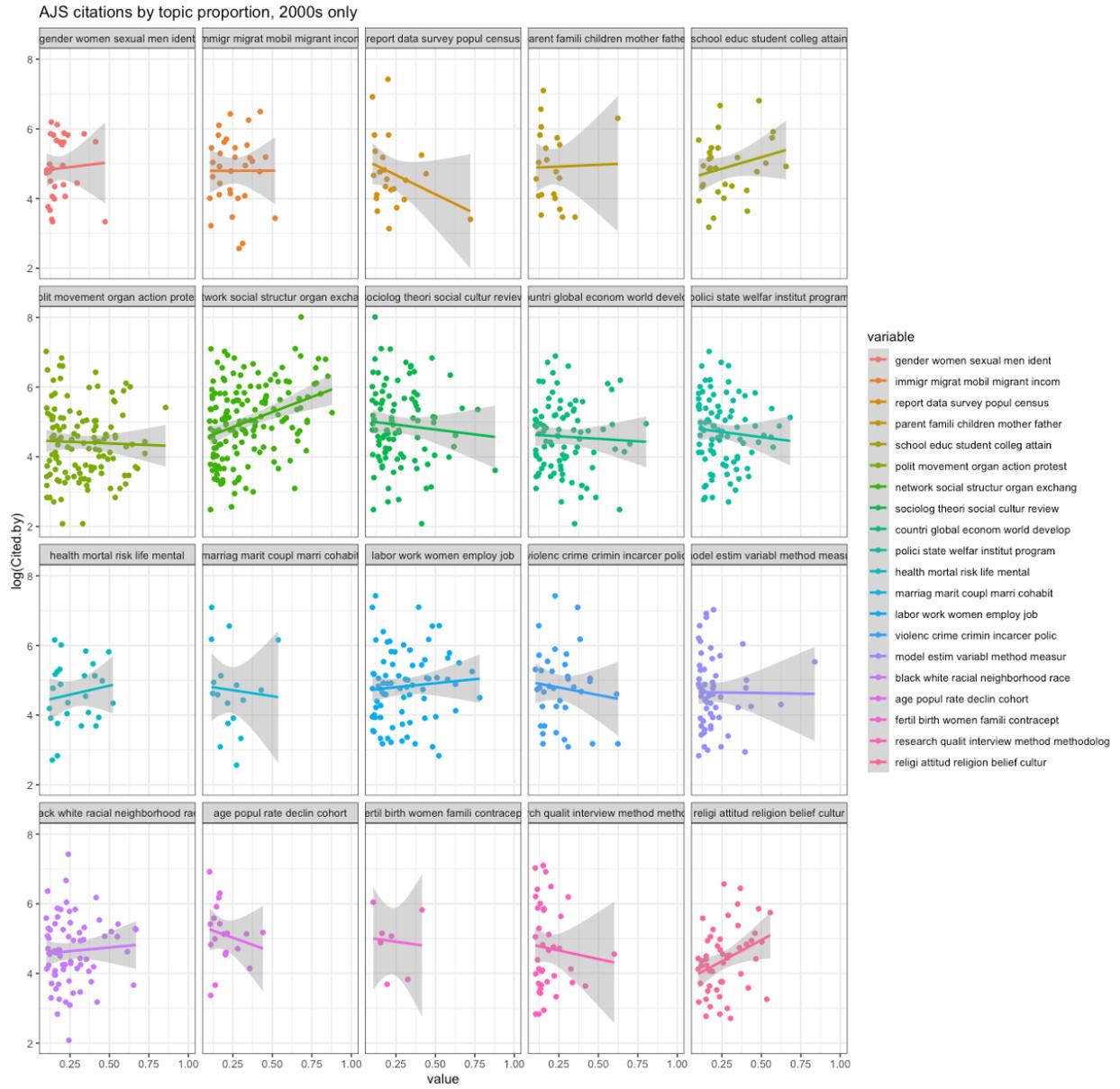

Figure S6: Incoming citations by topic proportion

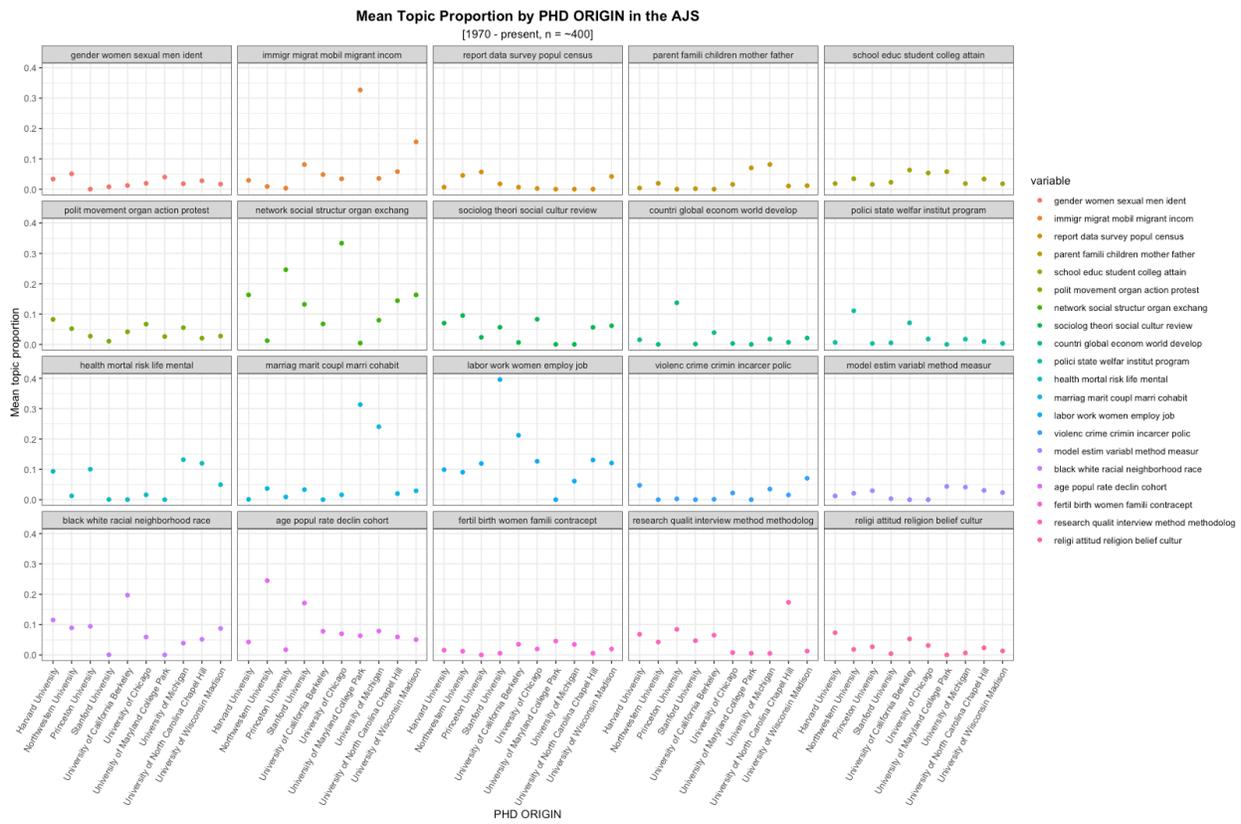

Figure S7: Mean topic proportion by PhD origin

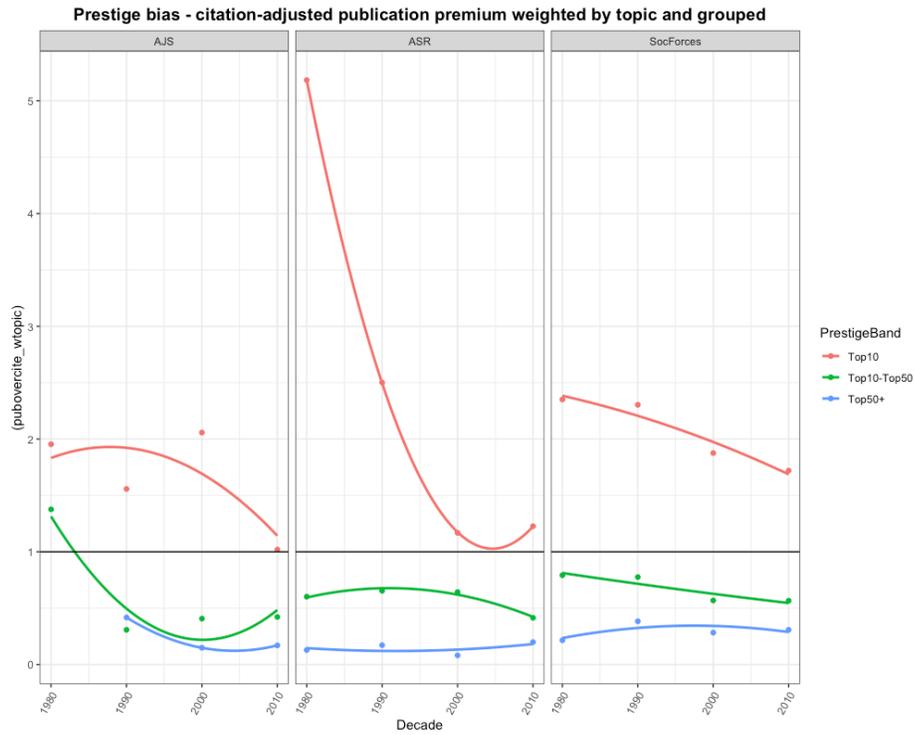

Figure S8: Prestige bias weighted by topic

A replication of Figure 3 with a topic-specific citation multiple and publication multiple. This suggests that our work is robust to PhD alum from particular departments specializing preferentially in different topics.

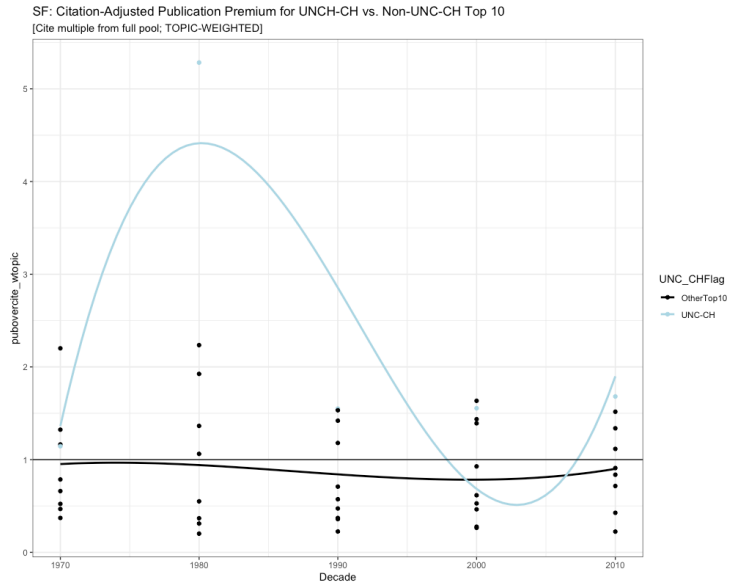

Figure S9: House bias in SF weighted by topic

A replication of Figure 4 with a topic-specific citation multiple and publication multiple. This suggests that our work is robust to PhD alum from particular departments specializing preferentially in different topics.

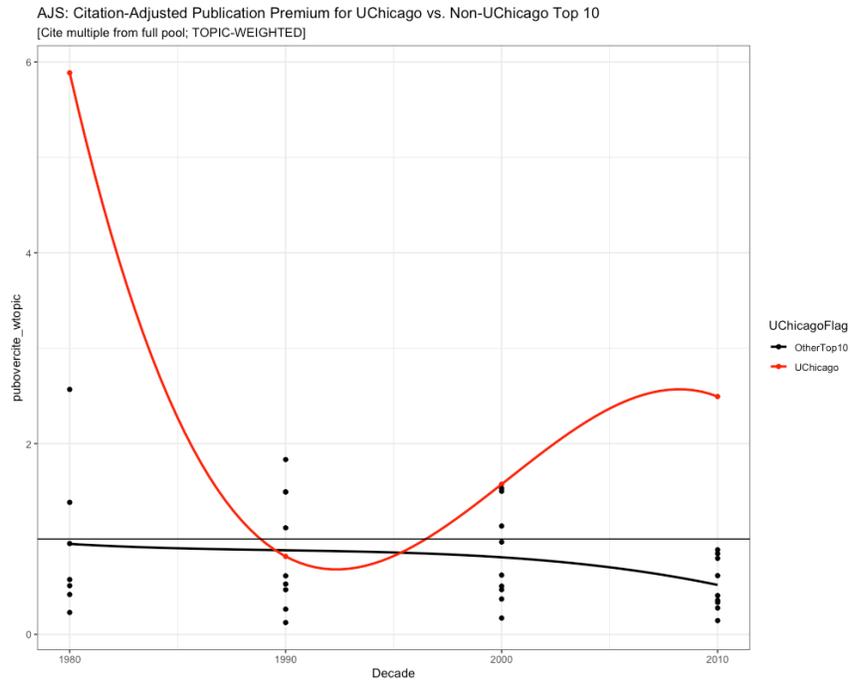

Figure S10: House bias in the AJS weighted by topic

A replication of Figure 5 with a topic-specific citation multiple and publication multiple. This suggests that our work is robust to PhD alum from particular departments specializing preferentially in different topics.